# Secure Beamforming in Full-Duplex SWIPT Systems With Loopback Self-Interference Cancellation

Yanjie Dong‡, Ahmed El Shafie§, Md. Jahangir Hossain†, Julian Cheng†, Naofal Al-Dhahir§, and Victor C. M. Leung‡

‡Department of Electrical and Computer Engineering, The University of British Columbia, Vancouver, BC, Canada
§Department of Electrical Engineering, The University of Texas at Dallas, Dallas, TX, USA
†School of Engineering, The University of British Columbia, Kelowna, BC, Canada

***Abstract*—Security is a critical issue in full duplex (FD) communication systems due to the broadcast nature of wireless channels. In this paper, joint design of information and artificial noise beamforming vectors is proposed for the FD simultaneous wireless information and power transferring (FD-SWIPT) systems with loopback self-interference cancellation. To guarantee high security and energy harvesting performance of the FD-SWIPT system, the proposed design is formulated as a secrecy rate maximization problem under energy transfer rate constraints. Although the secrecy rate maximization problem is non-convex, we solve it via semidefinite relaxation and a two-dimensional search. We prove the optimality of our proposed algorithm and demonstrate its performance via simulations.**

***Index Terms*—Beamforming, coordinated jamming, full-duplex communications, physical layer security, SWIPT.**

## I. Introduction

The upsurging wireless data traffic drives the industrial and academic communities to develop efficient ways to boost the usage of already scarce radio-frequency (RF) spectrum in the fifth generation (5G) wireless communication systems [1]. Full-duplex (FD) technology is considered as one of the promising solutions due to its ability to almost double the spectrum efficiency. However, FD nodes suffer from a strong loopback self-interference (LSI) due to the short distance between the transmitter and receiver sides on an FD node [1]. Several LSI cancellation techniques are proposed in the FD literature, e.g., digital-analog domain cancellation [2] and spatial domain suppression [3], [4]. Due to dynamic range limitations of the digital-analog domain cancellation reception circuits, the spatial domain suppression, which utilizes the spatial diversity to suppress the LSI, is required before further processing of the received signals [3], [4].

Simultaneous wireless information and power transferring (SWIPT) systems have received increased attention recently [5]–[7]. Based on the energy carried by the RF signals, SWIPT systems can deliver information signal to the user equipments (UEs) and energy signal to the low energy-consumption energy receivers (ERs) [6]. Hence, the integration of FD technology into the SWIPT systems is promising due to the simultaneous support of UEs and ERs, and the improvement of the spectrum efficiency and the energy efficiency [8]. To avoid the attenuation of the energy signal and meet energy demands, the ERs are usually deployed near the transmitters [9]. However, the strong received signal strength at the ERs can also benefit eavesdropping when the ERs are malicious users. Therefore, there exists a critical tradeoff between the security and energy requirements at the ERs in FD-SWIPT systems. It has been demonstrated in [10] that by using part of the transmission power to transmit artificial noise (AN), the information leakage to the eavesdroppers is degraded. Hence, the security in wireless communications is improved via broadcasting the AN [10]–[13]. Motivated by this fact, we assume that information and AN beamforming vectors are jointly generated by the FD base station (FD-BST) and FD user equipment (FD-UE) to suppress the information leakage to the ERs. Another benefit of this setup is that the FD-BST and FD-UE can coordinately perform AN beamforming, which improves the suppression efficiency of the information leakage [10]–[14].

Resource allocation is one of the key mechanisms to ensure communication security in wireless communication systems, such as FD communication systems and SWIPT systems. On the one hand, secure resource allocation in FD communication systems has been widely investigated for various topologies, e.g., point-to-point (PtP) topology [11], [12], [15], [16], point-to-multipoint (PtMP) topology [14], and wireless relaying topology [4], [11], [17]. Two major optimization problems are considered in secure FD communication systems, namely, secrecy-capacity maximization [11], [12], [16], [17] and transmission-power minimization [14]. For example, using jamming signals by the legitimate receiver, the authors in [11] studied secrecy-capacity maximization in FD communication systems with a PtP topology. Using the secrecy-capacity region in [18], [19], the authors in [16] proposed an alternating optimization algorithm to obtain the locally optimal solution to maximize the secrecy capacity of the secure FD communication systems with a PtP topology.

On the other hand, secure resource allocation in the SWIPT systems was investigated in [8], [9], [20]–[25]. To guarantee system security, different performance metrics, such as secrecy capacity [8], [9], [21], [24] and transmission power [20], [25], are optimized. In SWIPT with a single legitimate UE, the authors in [9] and [21] investigated secrecy-capacity maximization for multiple single-antenna eavesdroppers and one single multiple-antenna eavesdropper, respectively. The authors in [25] investigated the power-minimization problem in SWIPT systems with multiple ERs and multiple multi-antenna eavesdroppers under the imperfect channel state information (CSI). In addition, the authors in [22] analyzed the secrecy capacity in SWIPT systems with a single legitimate UE and multiple eavesdroppers, and they obtained closed-form expressions for the security outage probability and the ergodic security capacity. The authors in [8] studied weighted-capacity maximization in the wireless relaying system, where the BST operates in the FD mode and a single ER can envisage energy and eavesdrop information from the received signals. Despite the plethora of research works on the security of wireless communication systems [8], [9], [11]–[16], [20]–[25], they are mainly based on the assumption that each eavesdropper can acquire information from one legitimate UE. Hence, the reported secrecy rates are, generally, not achievable when the ER can eavesdrop information from the simultaneous UE and BS transmissions [18], [19].

Motivated by the pioneering work on the capacity-region of FD two-way communication systems [18], [19], we consider a more general scenario, where there are multiple ERs in the FD-SWIPT system. In the FD-SWIPT system, the bidirectional communications between FD-BST and FD-UE place requirements on communication security and harvested energy at the multiple ERs. To the authors' best knowledge, the joint design of the information beamforming and AN vectors in the FD-SWIPT system with multiple ERs has not been reported in the current literature. In this work, our objective is to maximize the secrecy rate of the FD-SWIPT system with LSI cancellation.

Our main contributions are summarized as follows

- To avoid the saturation of the digital and analog circuits caused by the strong LSI, we jointly design the information and AN vectors to cancel the LSI such that a certain secrecy rate is achieved.
- We investigate the secrecy rate maximization (SRM) problem in FD-SWIPT systems with LSI cancellation.
- We propose a two-stage algorithm to solve the formulated SRM problem. Numerical results are used to demonstrate the performance of our proposed algorithm.

*Notations:* Vectors and matrices are shown in bold lowercase letters and bold uppercase letters, respectively. $\mathbb{C}$ denotes the set of complex numbers. $\sim$ stands for "distributed as". $\boldsymbol{I}$ denotes an identity matrix, and $\mathbf{0}_{M\times N}$ denotes a zero matrix with $M$ rows and $N$ columns. The expectation of a random variable is denoted as $\mathbb{E}[\cdot]$. vec $[\cdot]$ converts an $M\times N$ matrix into a column vector of size $MN\times 1$. $\{\boldsymbol{w}_n\}_{n\in\mathcal{N}}$ represents the set composed of $\boldsymbol{w}_n$, $n\in\mathcal{N}$. For a square matrix $\boldsymbol{W}$, $\boldsymbol{W}^{\mathrm{H}}$ and $\mathrm{Tr}(\boldsymbol{W})$ denote its conjugate transpose and trace, respectively. $\boldsymbol{W}\succeq\mathbf{0}$ and $\boldsymbol{W}\succ\mathbf{0}$ imply that $\boldsymbol{W}$ is a positive semidefinite and a positive definite matrix, respectively.

## II. System Model and Problem Formulation

### A. System Model and Assumptions

We consider an FD-SWIPT system which consists of an FD-BST, a legitimate FD-UE, and $K$ ERs. Let $\mathcal{K}=\{1,2,\ldots,K\}$ be the set of ERs, where each ER can also be the potential eavesdroppers. The FD-BST is equipped with $N_a>1$ transmit antennas and a single receive antenna for simultaneous transmission and reception over the same frequency band. Similarly, the legitimate FD-UE is equipped with $N_b>1$ transmit antennas and a single receive antenna. Thus, the FD-BST and the FD-UE operate in FD mode. The $k$-th ER is equipped with a single receive antenna. We consider frame-based communication over frequency-nonselective fading channels with unit duration for each frame. Therefore, the terms "power" and "energy" can be used interchangeably.

We assume that the FD-BST and the FD-UE can perfectly estimate the CSI [4]. This assumption is reasonable since the ERs and the FD-UE are assumed to be non-hostile legitimate nodes in the FD-SWIPT system. However, the ERs are curious to decode the information signals transmitted by the FD-BST and the FD-UE. Therefore, the ERs are assumed to be the potential eavesdroppers. At the beginning of each frame, each node sends a pilot signal to the FD-BST. All nodes listen to the pilot signals and estimate the channels associated with the node sending the pilot signal. After that, the FD-BST sends another pilot signal such that each node estimates the CSI connected to the FD-BST. Then, the FD-UE feeds back the estimated CSI to the FD-BST. Finally, the FD-BST and the FD-UE use the beamforming vectors for information transmission and the AN signals for coordinated jamming.

The transmission signals of the FD-BST and FD-UE are, respectively, given by $\boldsymbol{x}_{a,b}=\boldsymbol{w}_{a,b}s_b+\boldsymbol{v}_a$ and $\boldsymbol{x}_{b,a}=\boldsymbol{w}_{b,a}s_a+\boldsymbol{v}_b$, where $s_a\sim\mathcal{CN}(0,1)$ and $s_b\sim\mathcal{CN}(0,1)$ denote, respectively, the information-bearing signals for the FD-BST and the FD-UE, and $s_a$ and $s_b$ are independent from each other. The vectors $\boldsymbol{w}_{a,b}\in\mathbb{C}^{N_a\times 1}$ and $\boldsymbol{v}_a\in\mathbb{C}^{N_a\times 1}$ ($\boldsymbol{w}_{b,a}\in\mathbb{C}^{N_b\times 1}$ and $\boldsymbol{v}_b\in\mathbb{C}^{N_b\times 1}$) are, respectively, the information beamforming vector and the AN vector of the FD-BST (FD-UE). In particular, the AN vector $\boldsymbol{v}_a$ ($\boldsymbol{v}_b$) is modeled as a circularly symmetric complex Gaussian vector with mean zero and covariance $\boldsymbol{V}_{a,a}\succeq 0$ ($\boldsymbol{V}_{b,b}\succeq 0$).

Therefore, the received signal at the $k$-th ER is given by $y_{e_k}=\boldsymbol{h}^{\mathrm{H}}_{a,e_k}\boldsymbol{w}_{a,b}s_b+\boldsymbol{h}^{\mathrm{H}}_{b,e_k}\boldsymbol{w}_{b,a}s_a+\boldsymbol{h}^{\mathrm{H}}_{e_k}\boldsymbol{v}+z_{e_k}$, where $\boldsymbol{h}_{e_k}\triangleq\mathrm{vec}\left([\boldsymbol{h}_{a,e_k},\boldsymbol{h}_{b,e_k}]\right)$ with the channel coefficient vectors from the FD-BST and the FD-UE to the $k$-th ER denoted by $\boldsymbol{h}_{a,e_k}\in\mathbb{C}^{N_a\times 1}$ and $\boldsymbol{h}_{b,e_k}\in\mathbb{C}^{N_b\times 1}$, respectively. The term $\boldsymbol{z}_{e_k}\sim\mathcal{CN}\left(0,\sigma^2_{e_k}\right)$ denotes the additive white Gaussian noise (AWGN) at the $k$-th ER with zero mean and variance $\sigma^2_{e_k}$; $\boldsymbol{v}\triangleq\mathrm{vec}\left([\boldsymbol{v}_a,\boldsymbol{v}_b]\right)$ is the compact form of the coordinated AN vector[1]. Here, the AN vector $\boldsymbol{v}$ is modeled as a circularly symmetric Gaussian vector with zero mean and covariance matrix $\boldsymbol{V}$. The covariance matrix $\boldsymbol{V}$ is given by

$$\boldsymbol{V}=\begin{bmatrix}\boldsymbol{V}_{a,a} & \boldsymbol{V}_{a,b}\\ \boldsymbol{V}_{b,a} & \boldsymbol{V}_{b,b}\end{bmatrix} \tag{1}$$

where the matrices $\boldsymbol{V}_{a,b}$ and $\boldsymbol{V}_{b,a}$ are defined as $\boldsymbol{V}_{a,b}=\mathbb{E}\left[\boldsymbol{v}_a\boldsymbol{v}_b^{\mathrm{H}}\right]$ and $\boldsymbol{V}_{b,a}=\mathbb{E}\left[\boldsymbol{v}_b\boldsymbol{v}_a^{\mathrm{H}}\right]$, respectively.

The achievable sum data rate per unit bandwidth of the $k$-th ER is given by

$$\begin{aligned}&C_{e_k}\left(\boldsymbol{W}_{a,b},\boldsymbol{W}_{b,a},\boldsymbol{V}\right)\\ &=\log\left(1+\frac{\mathrm{Tr}\left(\boldsymbol{H}_{a,e_k}\boldsymbol{W}_{a,b}\right)+\mathrm{Tr}\left(\boldsymbol{H}_{b,e_k}\boldsymbol{W}_{b,a}\right)}{\mathrm{Tr}\left(\boldsymbol{H}_{e_k}\boldsymbol{V}\right)+\sigma^2_{e_k}}\right).\end{aligned} \tag{2}$$

In addition, the following two achievable data rates per unit bandwidth of the $k$-th ER are, respectively, given by

$$\begin{aligned}&C_{a,e_k}\left(\boldsymbol{W}_{a,b},\boldsymbol{W}_{b,a},\boldsymbol{V}\right)\\ &=\log\left(1+\frac{\mathrm{Tr}\left(\boldsymbol{H}_{b,e_k}\boldsymbol{W}_{b,a}\right)}{\mathrm{Tr}\left(\boldsymbol{H}_{a,e_k}\boldsymbol{W}_{a,b}\right)+\mathrm{Tr}\left(\boldsymbol{H}_{e_k}\boldsymbol{V}\right)+\sigma^2_{e_k}}\right)\end{aligned} \tag{3}$$

and

$$\begin{aligned}&C_{b,e_k}\left(\boldsymbol{W}_{a,b},\boldsymbol{W}_{b,a},\boldsymbol{V}\right)\\ &=\log\left(1+\frac{\mathrm{Tr}\left(\boldsymbol{H}_{a,e_k}\boldsymbol{W}_{a,b}\right)}{\mathrm{Tr}\left(\boldsymbol{H}_{b,e_k}\boldsymbol{W}_{b,a}\right)+\mathrm{Tr}\left(\boldsymbol{H}_{e_k}\boldsymbol{V}\right)+\sigma^2_{e_k}}\right)\end{aligned} \tag{4}$$

[1] Using the channel matrices between the legitimate nodes, the AN symbols are generated from a pseudo-random signal which is perfectly known at the legitimate nodes but not at the eavesdroppers. This is realized by using a short secret key as a seed for the Gaussian pseudo-random sequence generator. The legitimate nodes regularly change the secret key seeds to maintain the AN sequence secured from the eavesdropper. A similar assumption can be found in, e.g., [15], [26] and the references therein.

where $\boldsymbol{W}_{a,b} \triangleq \boldsymbol{w}_{a,b}\boldsymbol{w}_{a,b}^{\mathrm{H}}$, $\boldsymbol{W}_{b,a} \triangleq \boldsymbol{w}_{b,a}\boldsymbol{w}_{b,a}^{\mathrm{H}}$, $\boldsymbol{H}_{a,e_k} \triangleq \boldsymbol{h}_{a,e_k}\boldsymbol{h}_{a,e_k}^{\mathrm{H}}$, $\boldsymbol{H}_{b,e_k} \triangleq \boldsymbol{h}_{b,e_k}\boldsymbol{h}_{b,e_k}^{\mathrm{H}}$ and $\boldsymbol{H}_{e_k} \triangleq \boldsymbol{h}_{e_k}\boldsymbol{h}_{e_k}^{\mathrm{H}}$. The ranks of beamforming matrices $\boldsymbol{W}_{a,b}$ and $\boldsymbol{W}_{b,a}$ are upper-bounded as $\mathrm{Rank}\left(\boldsymbol{W}_{a,b}\right) \leq 1$ and $\mathrm{Rank}\left(\boldsymbol{W}_{b,a}\right) \leq 1$.

The amount of harvested energy at the $k$-th ER is given by

$$\begin{aligned} &E_k\left(\boldsymbol{W}_{a,b}, \boldsymbol{W}_{b,a}, \boldsymbol{V}\right) \\ =&\eta\left(\mathrm{Tr}\left(\boldsymbol{H}_{e_k}\boldsymbol{V}\right) + \mathrm{Tr}\left(\boldsymbol{H}_{a,e_k}\boldsymbol{W}_{a,b}\right) + \mathrm{Tr}\left(\boldsymbol{H}_{b,e_k}\boldsymbol{W}_{b,a}\right) + \sigma_{e_k}^2\right) \end{aligned} \tag{5}$$

where $0 \leq \eta \leq 1$ is the efficiency of the energy harvester.

In addition, the received signals at the FD-BST and the FD-UE are given by $y_a = \boldsymbol{h}_{b,a}^{\mathrm{H}}\boldsymbol{w}_{b,a}s_a + \boldsymbol{h}_{a,a}^{\mathrm{H}}\boldsymbol{w}_{a,b}s_b + \boldsymbol{h}_a^{\mathrm{H}}\boldsymbol{v} + z_a$ and $y_b = \boldsymbol{h}_{a,b}^{\mathrm{H}}\boldsymbol{w}_{a,b}s_b + \boldsymbol{h}_{b,b}^{\mathrm{H}}\boldsymbol{w}_{b,a}s_a + \boldsymbol{h}_b^{\mathrm{H}}\boldsymbol{v} + z_b$, where $z_a \sim \mathcal{CN}\left(0, \sigma_a^2\right)$ and $z_b \sim \mathcal{CN}\left(0, \sigma_b^2\right)$ denote, respectively, the AWGN at the FD-BST and FD-UE; $\boldsymbol{h}_a \triangleq \mathrm{vec}\left(\left[\boldsymbol{h}_{a,a}, \boldsymbol{h}_{b,a}\right]\right)$ and $\boldsymbol{h}_b \triangleq \mathrm{vec}\left(\left[\boldsymbol{h}_{a,b}, \boldsymbol{h}_{b,b}\right]\right)$ denote the compact form of the channel coefficient vectors for the FD-BST and the FD-UE. Without loss of generality, we assume that the channel coefficient vectors $\boldsymbol{h}_a$, $\boldsymbol{h}_b$ and $\boldsymbol{h}_{e_k}$ are statistically independent.

The information transmission rate per unit bandwidth of the FD-UE and the FD-BST are, respectively, given by

$$\begin{aligned} &C_a\left(\boldsymbol{W}_{a,b}, \boldsymbol{W}_{b,a}, \boldsymbol{V}\right) \\ &= \log\left(1 + \frac{\mathrm{Tr}\left(\boldsymbol{H}_{b,a}\boldsymbol{W}_{b,a}\right)}{\mathrm{Tr}\left(\boldsymbol{H}_{a,a}\boldsymbol{W}_{a,b}\right) + \mathrm{Tr}\left(\boldsymbol{H}_a\boldsymbol{V}\right) + \sigma_a^2}\right) \end{aligned} \tag{6}$$

and

$$\begin{aligned} &C_b\left(\boldsymbol{W}_{a,b}, \boldsymbol{W}_{b,a}, \boldsymbol{V}\right) \\ &= \log\left(1 + \frac{\mathrm{Tr}\left(\boldsymbol{H}_{a,b}\boldsymbol{W}_{a,b}\right)}{\mathrm{Tr}\left(\boldsymbol{H}_{b,b}\boldsymbol{W}_{b,a}\right) + \mathrm{Tr}\left(\boldsymbol{H}_b\boldsymbol{V}\right) + \sigma_b^2}\right) \end{aligned} \tag{7}$$

where $\boldsymbol{H}_a \triangleq \boldsymbol{h}_a\boldsymbol{h}_a^{\mathrm{H}}$, $\boldsymbol{H}_{a,a} \triangleq \boldsymbol{h}_{a,a}\boldsymbol{h}_{a,a}^{\mathrm{H}}$, $\boldsymbol{H}_{a,b} \triangleq \boldsymbol{h}_{a,b}\boldsymbol{h}_{a,b}^{\mathrm{H}}$, $\boldsymbol{H}_b \triangleq \boldsymbol{h}_b\boldsymbol{h}_b^{\mathrm{H}}$, $\boldsymbol{H}_{b,b} \triangleq \boldsymbol{h}_{b,b}\boldsymbol{h}_{b,b}^{\mathrm{H}}$, and $\boldsymbol{H}_{b,a} \triangleq \boldsymbol{h}_{b,a}\boldsymbol{h}_{b,a}^{\mathrm{H}}$. Note that the signals received by the FD-BST and FD-UE contain the confidential information and the auxiliary information [19].

Based on the Corollary 1 in [19], the secrecy rate of the FD-SWIPT system is obtained as

$$\begin{aligned} &C^{\mathrm{SEC}}\left(\boldsymbol{W}_{a,b}, \boldsymbol{W}_{b,a}, \boldsymbol{V}\right) \\ =&C_a + C_b - \max\left\{\max_{k\in\mathcal{K}} C_{a,e_k} + \max_{k\in\mathcal{K}} C_{b,e_k}, \max_{k\in\mathcal{K}} C_{e_k}\right\} \end{aligned} \tag{8}$$

where $C_a$, $C_b$, $C_{a,e_k}$, $C_{b,e_k}$ and $C_{e_k}$ are, respectively, defined in (2), (3), (4), (6) and (7).

***Remark 1:*** From (8), we observe that the secrecy rate is related to the information transmission rates $C_a\left(\boldsymbol{W}_{a,b}, \boldsymbol{W}_{b,a}, \boldsymbol{V}\right)$ and $C_b\left(\boldsymbol{W}_{a,b}, \boldsymbol{W}_{b,a}, \boldsymbol{V}\right)$ as well as the information leakage rates $C_{e_k}\left(\boldsymbol{W}_{a,b}, \boldsymbol{W}_{b,a}, \boldsymbol{V}\right)$, $C_{a,e_k}\left(\boldsymbol{W}_{a,b}, \boldsymbol{W}_{b,a}, \boldsymbol{V}\right)$ and $C_{b,e_k}\left(\boldsymbol{W}_{a,b}, \boldsymbol{W}_{b,a}, \boldsymbol{V}\right)$, which indicates that the achievable rates at all nodes and the harvested energy at the ERs depend on the information and AN beamforming vectors. One of the major challenges in the application of FD technology is the strong LSI. In this work, we leverage the spatial domain technique to cancel the LSI [3]–[5].

### *B. Problem Formulation*

Our objective is to maximize the secrecy rate of the FD-SWIPT system via joint design of information and AN beamforming vectors subjected to the LSI cancellation constraints, the energy harvesting constraints and the maximum transmission power constraints. Therefore, the SRM problem is formulated as

$$\max_{\boldsymbol{W}_{a,b}, \boldsymbol{W}_{b,a}, \boldsymbol{V}} \; C^{\mathrm{SEC}}\left(\boldsymbol{W}_{a,b}, \boldsymbol{W}_{b,a}, \boldsymbol{V}\right) \tag{9a}$$

$$\text{s.t. } \mathrm{Tr}\left(\boldsymbol{B}_a\boldsymbol{V}\right) + \mathrm{Tr}\left(\boldsymbol{W}_{a,b}\right) \leq P_a^{\max} \tag{9b}$$

$$\mathrm{Tr}\left(\boldsymbol{B}_b\boldsymbol{V}\right) + \mathrm{Tr}\left(\boldsymbol{W}_{b,a}\right) \leq P_b^{\max} \tag{9c}$$

$$\begin{aligned} &\mathrm{Tr}\left(\boldsymbol{H}_{a,e_k}\boldsymbol{W}_{a,b}\right) + \mathrm{Tr}\left(\boldsymbol{H}_{b,e_k}\boldsymbol{W}_{b,a}\right) \\ &\quad + \mathrm{Tr}\left(\boldsymbol{H}_{e_k}\boldsymbol{V}\right) + \sigma_{e_k}^2 \geq \frac{P_k^{\mathrm{REQ}}}{\eta}, \forall k \end{aligned} \tag{9d}$$

$$\mathrm{Tr}\left(\boldsymbol{H}_a\boldsymbol{V}\right) = \mathrm{Tr}\left(\boldsymbol{H}_b\boldsymbol{V}\right) = 0 \tag{9e}$$

$$\mathrm{Tr}\left(\boldsymbol{H}_{a,a}\boldsymbol{W}_{a,b}\right) = \mathrm{Tr}\left(\boldsymbol{H}_{b,b}\boldsymbol{W}_{b,a}\right) = 0 \tag{9f}$$

$$\boldsymbol{W}_{a,b} \succeq 0, \boldsymbol{W}_{b,a} \succeq 0, \boldsymbol{V} \succeq 0 \tag{9g}$$

$$\mathrm{Rank}\left(\boldsymbol{W}_{a,b}\right) \leq 1 \text{ and } \mathrm{Rank}\left(\boldsymbol{W}_{b,a}\right) \leq 1 \tag{9h}$$

where $P_a^{\max}$ and $P_b^{\max}$ are maximum transmission power for the FD-BST and the FD-UE, respectively. $P_k^{\mathrm{REQ}}$ denotes the required amount of harvested energy by the $k$-th ER. Here, the matrices $\boldsymbol{B}_a$ and $\boldsymbol{B}_b$ are, respectively, defined as

$$\boldsymbol{B}_a = \begin{bmatrix} \boldsymbol{I} & \boldsymbol{0}_{N_a \times N_b} \\ \boldsymbol{0}_{N_b \times N_a} & \boldsymbol{0}_{N_b \times N_b} \end{bmatrix} \text{ and } \boldsymbol{B}_b = \begin{bmatrix} \boldsymbol{0}_{N_a \times N_a} & \boldsymbol{0}_{N_a \times N_b} \\ \boldsymbol{0}_{N_b \times N_a} & \boldsymbol{I} \end{bmatrix}. \tag{10}$$

## III. SRM With LSI Cancellation

### *A. Feasibility Analysis*

The SRM problem can be infeasible under certain channel conditions when the values of $P_a^{\max}$ and $P_b^{\max}$ are too low and/or the value of $P_k^{\mathrm{REQ}}$ is too high, $k \in \mathcal{K}$. In the case that the FD-BST and FD-UE do not transmit information, the consumed transmission power will be lowest. When the maximum transmission power of FD-BST and FD-UE cannot satisfy the energy requirements of the ERs, the SRM problem (9) is infeasible. Hence, the feasibility of the SRM optimization problem can be checked by setting the information beamforming matrices $\boldsymbol{W}_{a,b}$ and $\boldsymbol{W}_{b,a}$ to zero and solving the following problem

$$\begin{aligned} &\text{Find } \boldsymbol{V} \succeq \boldsymbol{0} \\ &\text{s.t. } \mathrm{Tr}\left(\boldsymbol{B}_a\boldsymbol{V}\right) \leq P_a^{\max} \\ &\quad \mathrm{Tr}\left(\boldsymbol{B}_b\boldsymbol{V}\right) \leq P_b^{\max} \\ &\quad \mathrm{Tr}\left(\boldsymbol{H}_{e_k}\boldsymbol{V}\right) + \sigma_{e_k}^2 \geq \frac{P_k^{\mathrm{REQ}}}{\eta}, \forall k. \end{aligned} \tag{11}$$

The optimization problem (11) is convex; therefore, we can check the feasibility via CVX [27]. If the matrix $\boldsymbol{V}$ is obtained via (11), we have $\left(\boldsymbol{W}_{a,b} = \boldsymbol{0}, \boldsymbol{W}_{b,a} = \boldsymbol{0}, \boldsymbol{V}\right)$ is a feasible solution to the SRM problem (9). Without loss of generality, we assume the SRM problem (9) is feasible hereinafter.

### *B. Joint Design of Beamforming and AN Vectors*

We observe that the SRM problem (9) is non-convex due to the non-convex objective function (9a) and the rank-one constraints in (9g). In order to develop practical algorithm to solve the SRM problem, we introduce two parameters $\theta \in [0, 1]$ and $t$ such that the SRM problem is recast into a parametric SRM (P-SRM) problem as

$$\max_{\boldsymbol{W}_{a,b},\boldsymbol{W}_{b,a},\boldsymbol{V},\theta,t} \mathcal{OBJ}\left(\boldsymbol{W}_{b,a},\boldsymbol{W}_{a,b}\right) \tag{12a}$$

$$\text{s.t. } C_{e_k}\left(\boldsymbol{W}_{a,b},\boldsymbol{W}_{b,a},\boldsymbol{V}\right) \leq t, \forall k \tag{12b}$$

$$C_{a,e_k}\left(\boldsymbol{W}_{a,b},\boldsymbol{W}_{b,a},\boldsymbol{V}\right) \leq \theta t, \forall k \tag{12c}$$

$$C_{b,e_k}\left(\boldsymbol{W}_{a,b},\boldsymbol{W}_{b,a},\boldsymbol{V}\right) \leq (1-\theta)t, \forall k \tag{12d}$$

$$\theta \in [0,1], t \geq 0 \tag{12e}$$

$$\text{(9b)} - \text{(9h)} \tag{12f}$$

where $\mathcal{OBJ}\left(\boldsymbol{W}_{b,a},\boldsymbol{W}_{a,b}\right) = \log\left(1+\mathrm{Tr}\left(\boldsymbol{H}_{b,a}\boldsymbol{W}_{b,a}\right)/\sigma_a^2\right) + \log\left(1+\mathrm{Tr}\left(\boldsymbol{H}_{a,b}\boldsymbol{W}_{a,b}\right)/\sigma_b^2\right) - t$.

With the introduced parameters, the objective function (12a) is convex. Moreover, the constraints in (12b)-(12d) are convex for given a parameter pair $(\theta, t)$. In order to obtain the optimal solution to (12), we develop a two-stage algorithm by: 1) solving the P-SRM problem under given $(\theta, t)$; 2) performing a two-dimensional search for the optimal $(\theta^*, t^*)$.

Next, we investigate the algorithm to solve the P-SRM problem with fixed parameters $\theta$ and $t$. Since the covariance matrix of AN do not interfere the communication between the FD-BST and the FD-UE from the constraints in (9e), we conclude that the covariance matrix $\boldsymbol{V}$ lies in the null space of the matrix $[\boldsymbol{h}_a, \boldsymbol{h}_b]^{\mathrm{H}}$. Let $\overline{\boldsymbol{Y}}$ denote the null space of the matrix $[\boldsymbol{h}_a, \boldsymbol{h}_b]^{\mathrm{H}}$, where $\overline{\boldsymbol{Y}}$ can be obtained via the singular value decomposition of the matrix $[\boldsymbol{h}_a, \boldsymbol{h}_b]^{\mathrm{H}}$ with $\overline{\boldsymbol{Y}}^{\mathrm{H}}\overline{\boldsymbol{Y}} = \boldsymbol{I}$. Hence, the AN vectors can be denoted as $\boldsymbol{v}_n = \overline{\boldsymbol{Y}}\overline{\boldsymbol{v}}_n$, $n = 1, 2, \ldots, N_a + N_b - 2$, where $\overline{\boldsymbol{v}}_n \in \mathbb{C}^{(N_a+N_b-2)\times 1}$, and the covariance matrix $\boldsymbol{V}$ is obtained as

$$\boldsymbol{V} = \overline{\boldsymbol{Y}}\,\overline{\boldsymbol{V}}\,\overline{\boldsymbol{Y}}^{\mathrm{H}}. \tag{13}$$

where $\overline{\boldsymbol{V}} = \sum_{n=1}^{N_a+N_b-2} \overline{\boldsymbol{v}}_n \overline{\boldsymbol{v}}_n^{\mathrm{H}}$.

Following similar arguments, we denote that the information beamforming vectors of the FD-BST and the FD-UE as $\boldsymbol{w}_{a,b} = \overline{\boldsymbol{X}}_{a,b}\overline{\boldsymbol{w}}_{a,b}$ and $\boldsymbol{w}_{b,a} = \overline{\boldsymbol{X}}_{b,a}\overline{\boldsymbol{w}}_{b,a}$, where $\overline{\boldsymbol{w}}_{a,b} \in \mathbb{C}^{(N_a-1)\times 1}$ and $\overline{\boldsymbol{w}}_{b,a} \in \mathbb{C}^{(N_b-1)\times 1}$; the matrices $\overline{\boldsymbol{X}}_{a,b} \in \mathbb{C}^{N_a\times(N_a-1)}$ and $\overline{\boldsymbol{X}}_{b,a} \in \mathbb{C}^{N_b\times(N_b-1)}$ are the null spaces of the channel coefficient vector $\boldsymbol{h}_{a,a}$ and $\boldsymbol{h}_{b,b}$ with $\overline{\boldsymbol{X}}_{a,b}^{\mathrm{H}}\overline{\boldsymbol{X}}_{a,b} = \boldsymbol{I}$ and $\overline{\boldsymbol{X}}_{b,a}^{\mathrm{H}}\overline{\boldsymbol{X}}_{b,a} = \boldsymbol{I}$. Hence, the information beamforming matrices of the FD-BST and the FD-UE are given as

$$\boldsymbol{W}_{a,b} = \overline{\boldsymbol{X}}_{a,b}\overline{\boldsymbol{W}}_{a,b}\overline{\boldsymbol{X}}_{a,b}^{\mathrm{H}} \tag{14}$$

and

$$\boldsymbol{W}_{b,a} = \overline{\boldsymbol{X}}_{b,a}\overline{\boldsymbol{W}}_{b,a}\overline{\boldsymbol{X}}_{b,a}^{\mathrm{H}} \tag{15}$$

where $\overline{\boldsymbol{W}}_{a,b} = \overline{\boldsymbol{w}}_{a,b}\overline{\boldsymbol{w}}_{a,b}^{\mathrm{H}}$ and $\overline{\boldsymbol{W}}_{b,a} = \overline{\boldsymbol{w}}_{b,a}\overline{\boldsymbol{w}}_{b,a}^{\mathrm{H}}$.

Substituting (13)-(15) into the P-SRM problem, we obtain a simplified P-SRM (SP-SRM) problem (16) at the top of next page with $\overline{\boldsymbol{H}}_{a,b} \triangleq \overline{\boldsymbol{X}}_{a,b}^{\mathrm{H}}\boldsymbol{H}_{a,b}\overline{\boldsymbol{X}}_{a,b}$, $\overline{\boldsymbol{H}}_{b,a} \triangleq \overline{\boldsymbol{X}}_{b,a}^{\mathrm{H}}\boldsymbol{H}_{b,a}\overline{\boldsymbol{X}}_{b,a}$, $\overline{\boldsymbol{H}}_{a,e_k} \triangleq \overline{\boldsymbol{X}}_{a,b}^{\mathrm{H}}\boldsymbol{H}_{a,e_k}\overline{\boldsymbol{X}}_{a,b}$, $\overline{\boldsymbol{H}}_{b,e_k} \triangleq \overline{\boldsymbol{X}}_{b,a}^{\mathrm{H}}\boldsymbol{H}_{b,e_k}\overline{\boldsymbol{X}}_{b,a}$, $\overline{\boldsymbol{H}}_{e_k} \triangleq \overline{\boldsymbol{Y}}^{\mathrm{H}}\boldsymbol{H}_{e_k}\overline{\boldsymbol{Y}}$, $\overline{\boldsymbol{B}}_a \triangleq \overline{\boldsymbol{Y}}^{\mathrm{H}}\boldsymbol{B}_a\overline{\boldsymbol{Y}}$ and $\overline{\boldsymbol{B}}_b \triangleq \overline{\boldsymbol{Y}}^{\mathrm{H}}\boldsymbol{B}_b\overline{\boldsymbol{Y}}$.

For a given parameter pair $(\theta, t)$, performing the semi-definite relaxation (SDR) to the SP-SRM problem can obtain a convex optimization problem. Therefore, the SDR version of the SP-SRM problem can be solved via CVX [27]. Then, we are motivated to investigate the tightness of the SDR and obtain the following Proposition 1.

***Proposition 1:*** When the SP-SRM problem (16) is feasible and the given parameter pair $(\theta, t)$ is in the feasible region of SP-SRM problem, there always exists a solution $(\overline{\boldsymbol{W}}_{b,a}^*, \overline{\boldsymbol{W}}_{a,b}^*, \overline{\boldsymbol{V}}^*)$ to the SDR version of the SP-SRM problem satisfying the rank-one constraints $\mathrm{Rank}(\overline{\boldsymbol{W}}_{a,b}^*) \leq 1$ and $\mathrm{Rank}(\overline{\boldsymbol{W}}_{b,a}^*) \leq 1$.

*Proof:* See Appendix. □

From Proposition 1, we observe that the optimal solutions to the SDR version of the SP-SRM problem always satisfy the rank-one constraints in (16i). Hence, the SDR is tight, and we obtain the optimal solution to the SP-SRM problem via solving the SDR version of the SP-SRM problem. The procedure of the proposed two-stage joint information and AN beamforming algorithm is as follows.

1) Given the parameter pair $(\theta, t)$, the FD-BST solves the SDR version of the SP-SRM problem.
2) the FD-BST performs the two-dimensional search for the parameter pair $(\theta, t)$.
3) Repeat Steps 1) and 2) until the optimal parameter pair $(\theta^*, t^*)$ is obtained.
4) the FD-BST does the rank-one constraint recovery in (25)-(27) if obtained $\overline{\boldsymbol{W}}_{a,b}^*$ and $\overline{\boldsymbol{W}}_{b,a}^*$ are not rank-one.

## IV. Numerical Results

The simulation parameters are set as follows. In the FD-SWIPT system, there are one FD-BST, one FD-UE and 3 ERs. The FD-BST and FD-UE are, respectively, equipped with 3 and 3 antennas. The maximum transmission power of the FD-BST and FD-UE are 23 dBm and 20 dBm, respectively. Without loss of generality, we assume all the channels in the FD-SWIPT system are Rayleigh fadings. The pathloss of the BST-UE channel, BST-ER channels and UE-ER channels are set as $-35$ dB, $-20$ dB and $-20$ dB, respectively. Moreover, the LSI channels are assumed to experience Rayleigh fadings with average power attenuation as $-20$ dB. The required harvested energy of each ER is set as 2 dBm. The efficiency of the energy harvester is 0.5.

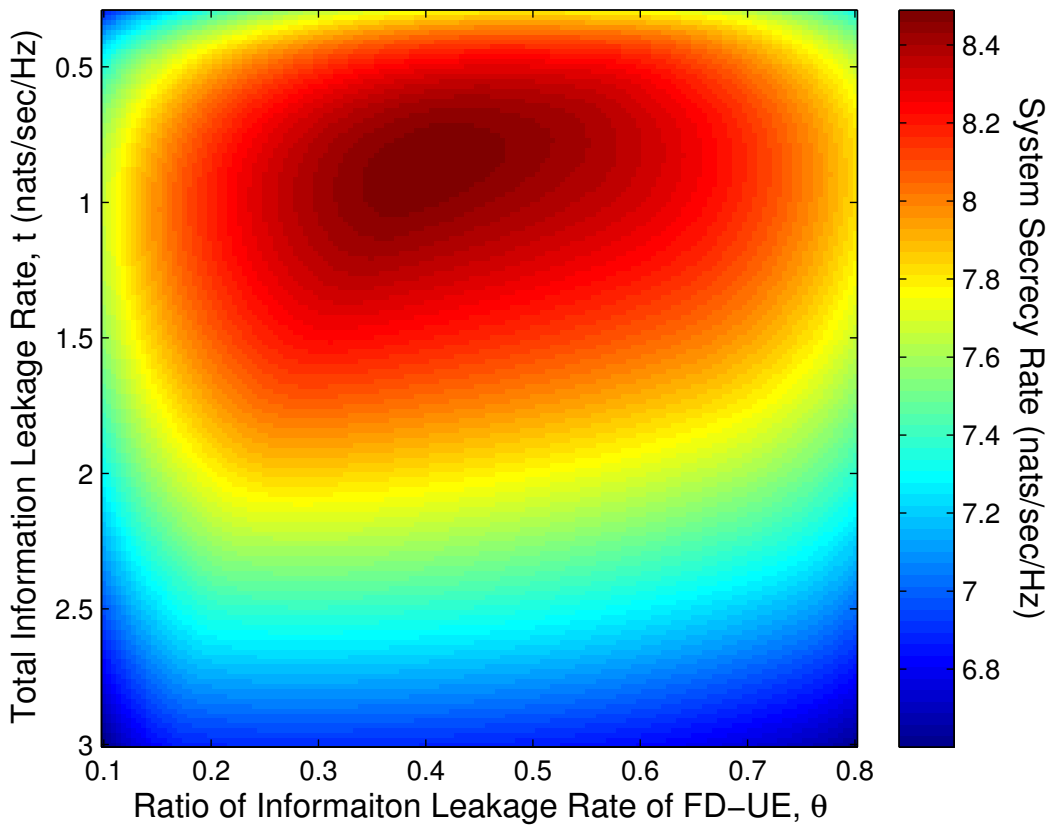

Fig. 1. An illustration of two-dimensional search for optimal secrecy rate.

Figure 1 shows the variation of the system secrecy rate over the introduced maximum information leakage rate $t$ and the ratio of the information leakage of the FD-UE $\theta$ based on one single channel implementation. With the color deepening from blue to red, the system secrecy rate increases in Fig. 1. We observe that the region with red color is a convex region. Therefore, the system secrecy rate is jointly concave

$$
\begin{aligned}
\max_{\overline{\boldsymbol{W}}_{b,a}, \overline{\boldsymbol{W}}_{a,b}, \overline{\boldsymbol{V}}, \theta, t} \quad & \log\left(1 + \frac{\mathrm{Tr}\left(\overline{\boldsymbol{H}}_{b,a}\overline{\boldsymbol{W}}_{b,a}\right)}{\sigma_a^2}\right) + \log\left(1 + \frac{\mathrm{Tr}\left(\overline{\boldsymbol{H}}_{a,b}\overline{\boldsymbol{W}}_{a,b}\right)}{\sigma_b^2}\right) - t && \text{(16a)}\\
\text{s.t.} \quad & \frac{1}{\exp(t) - 1}\left(\mathrm{Tr}\left(\overline{\boldsymbol{H}}_{a,e_k}\overline{\boldsymbol{W}}_{a,b}\right) + \mathrm{Tr}\left(\overline{\boldsymbol{H}}_{b,e_k}\overline{\boldsymbol{W}}_{b,a}\right)\right) \le \mathrm{Tr}\left(\overline{\boldsymbol{H}}_{e_k}\overline{\boldsymbol{V}}\right) + \sigma_{e_k}^2, \forall k && \text{(16b)}\\
& \frac{1}{\exp(\theta t) - 1}\mathrm{Tr}\left(\overline{\boldsymbol{H}}_{b,e_k}\overline{\boldsymbol{W}}_{b,a}\right) \le \mathrm{Tr}\left(\overline{\boldsymbol{H}}_{a,e_k}\overline{\boldsymbol{W}}_{a,b}\right) + \mathrm{Tr}\left(\overline{\boldsymbol{H}}_{e_k}\overline{\boldsymbol{V}}\right) + \sigma_{e_k}^2, \forall k && \text{(16c)}\\
& \frac{1}{\exp((1-\theta)t) - 1}\mathrm{Tr}\left(\overline{\boldsymbol{H}}_{a,e_k}\overline{\boldsymbol{W}}_{a,b}\right) \le \mathrm{Tr}\left(\overline{\boldsymbol{H}}_{b,e_k}\overline{\boldsymbol{W}}_{b,a}\right) + \mathrm{Tr}\left(\overline{\boldsymbol{H}}_{e_k}\overline{\boldsymbol{V}}\right) + \sigma_{e_k}^2, \forall k && \text{(16d)}\\
& \mathrm{Tr}\left(\overline{\boldsymbol{B}}_a\overline{\boldsymbol{V}}\right) + \mathrm{Tr}\left(\overline{\boldsymbol{W}}_{a,b}\right) \le P_a^{\max} && \text{(16e)}\\
& \mathrm{Tr}\left(\overline{\boldsymbol{B}}_b\overline{\boldsymbol{V}}\right) + \mathrm{Tr}\left(\overline{\boldsymbol{W}}_{b,a}\right) \le P_b^{\max} && \text{(16f)}\\
& \mathrm{Tr}\left(\overline{\boldsymbol{H}}_{a,e_k}\overline{\boldsymbol{W}}_{a,b}\right) + \mathrm{Tr}\left(\overline{\boldsymbol{H}}_{b,e_k}\overline{\boldsymbol{W}}_{b,a}\right) + \mathrm{Tr}\left(\overline{\boldsymbol{H}}_{e_k}\overline{\boldsymbol{V}}\right) + \sigma_{e_k}^2 \ge \frac{P_k^{\mathrm{REQ}}}{\eta}, \forall k && \text{(16g)}\\
& \overline{\boldsymbol{W}}_{a,b} \succeq \mathbf{0}, \overline{\boldsymbol{W}}_{b,a} \succeq \mathbf{0} \text{ and } \overline{\boldsymbol{V}} \succeq \mathbf{0} && \text{(16h)}\\
& \mathrm{Rank}\left(\overline{\boldsymbol{W}}_{a,b}\right) \le 1 \text{ and } \mathrm{Rank}\left(\overline{\boldsymbol{W}}_{b,a}\right) \le 1 && \text{(16i)}\\
& \theta \in [0,1] \text{ and } t \ge 0 && \text{(16j)}
\end{aligned}
$$

in maximum bound of information leakage rate $t$ and the the ratio of the maximum information leakage rate of the FD-BST $\theta$. We conclude that the optimal system secrecy rate can be obtained via our proposed two-stage joint information and AN beamforming algorithm. Moreover, the first search dimension is the information leakage rate, and the second search dimension is the ratio of the maximum information leakage rate of the FD-BST. This is due to the fact that the system secrecy rate is defined as the difference between the sum information transmission rate $C_a + C_b$ and the information leakage rate $\max\{\max_{k\in\mathcal{K}} C_{a,e_k} + \max_{k\in\mathcal{K}} C_{b,e_k}, \max_{k\in\mathcal{K}} C_{e_k}\}$.

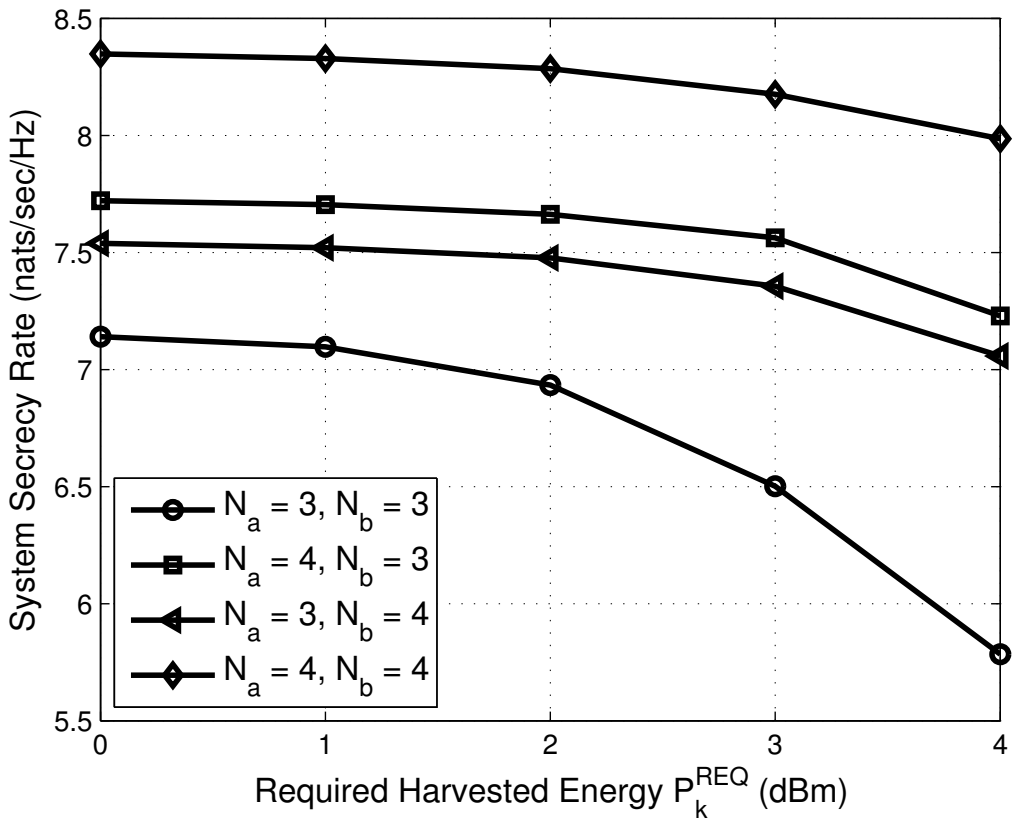

Fig. 2. The variation of the average secrecy rate of the FD-SWIPT system with the required harvested energy of each ER.

Figure 2 depicts the variation of the system secrecy rate of the FD-SWIPT system with the required harvested energy of each ER based on average of 200 implementations. We observe that the system secrecy rate decreases monotonically with increasing required harvested energy of each ER. When the number of transmit antennas increases from the $N_a = 3$ and $N_b = 3$, to $N_a = 3$ and $N_b = 4$, the decreasing speed of the system secrecy rate becomes smaller. Similar observations can also be made when the the number of transmit antennas increases from the $N_a = 3$ and $N_b = 4$, to $N_a = 4$ and $N_b = 4$. This is due to the fact that more transmit antenna enables the FD-SWIPT system to more degree of freedom to suppress the information leakage to the ERs. We also observe that the system secrecy rate in the case $N_a = 3$ and $N_b = 4$ is lower than the system secrecy rate in the case $N_a = 4$ and $N_b = 4$. This observation can be explained as follows. Under the same degree of freedom, increasing the maximum transmit power at the larger antenna node (the case $N_a = 4$ and $N_b = 4$) can suppress the information leakage more effectively than increasing the maximum transmit power at the lower antenna node (the case $N_a = 3$ and $N_b = 4$).

## V. Conclusions

We investigated the SRM problem under the constraints of LSI cancellation and required energy transfer rate in FD-SWIPT systems. Numerical results demonstrated that the optimal system secrecy rate can be obtained via the two-dimensional search for the SRM problem. We showed that there exists a tradeoff between the secrecy rate and energy transfer rate. When one of the two FD nodes has more antennas, increasing the maximum transmission power of this FD node can improve the secrecy rate than increasing the maximum transmission power of another FD node. Increasing the number of antennas on the FD nodes improves the secrecy rate of the FD-SWIPT system. Moreover, we showed that the achievable secrecy rate decreases as the energy transfer rate requirement increases, which is reasonable since the feasibility region shrinks and hence a lower objective value is expected. Finally, we showed that the impact of increasing the energy transfer rate requirement is increased when the number of antennas at the legitimate transmitters is low.

## Appendix

Introducing nonnegative dual variables $\mu$, $\nu$, $\{\lambda_k, \beta_k, \gamma_k, \varpi_k\}_{k\in\mathcal{K}}$ and positive semidefinite dual matrices $\boldsymbol{\Sigma}_{a,b}$, $\boldsymbol{\Sigma}_{b,a}$ and $\boldsymbol{\Upsilon}$, the Karush-Khun-Tucker conditions associated with the matrices $\overline{\boldsymbol{W}}_{b,a}$ and $\overline{\boldsymbol{W}}_{a,b}$ are, respectively, given as

$$\boldsymbol{C}_{b,a}^* - \frac{\overline{\boldsymbol{H}}_{b,a}}{\sigma_a^2 + \mathrm{Tr}\left(\overline{\boldsymbol{H}}_{b,a}\overline{\boldsymbol{W}}_{b,a}^*\right)} - \boldsymbol{\Sigma}_{b,a}^* = 0 \qquad (17)$$

$$\boldsymbol{\Sigma}_{b,a}^*\overline{\boldsymbol{W}}_{b,a}^* = 0 \qquad (18)$$

and

$$\boldsymbol{C}^*_{a,b} - \frac{\overline{\boldsymbol{H}}_{a,b}}{\sigma_b^2 + \text{Tr}\left(\overline{\boldsymbol{H}}_{a,b}\overline{\boldsymbol{W}}^*_{a,b}\right)} - \boldsymbol{\Sigma}^*_{a,b} = 0 \quad (19)$$

$$\boldsymbol{\Sigma}^*_{a,b}\overline{\boldsymbol{W}}^*_{a,b} = 0 \quad (20)$$

where

$$\boldsymbol{C}^*_{b,a} = \sum_{k=1}^{K}\left(\frac{\lambda_k^*}{e^t - 1} + \frac{\beta_k^*}{e^{\theta t} - 1} - \gamma_k^* - \varpi_k^*\right)\overline{\boldsymbol{H}}_{b,e_k} + \nu^*\boldsymbol{I} \quad (21)$$

and

$$\boldsymbol{C}^*_{a,b} = \sum_{k=1}^{K}\left(\frac{\lambda_k^*}{e^t - 1} - \beta_k^* + \frac{\gamma_k^*}{e^{(1-\theta)t} - 1} - \varpi_k^*\right)\overline{\boldsymbol{H}}_{a,e_k} + \mu^*\boldsymbol{I}. \quad (22)$$

Based on (17) and (18), we obtain

$$\boldsymbol{C}^*_{b,a}\overline{\boldsymbol{W}}^*_{b,a} = \frac{\overline{\boldsymbol{H}}_{b,a}\overline{\boldsymbol{W}}^*_{b,a}}{\sigma_a^2 + \text{Tr}\left(\overline{\boldsymbol{H}}_{b,a}\overline{\boldsymbol{W}}^*_{b,a}\right)}. \quad (23)$$

If the matrix $\boldsymbol{C}^*_{b,a}$ is full rank, we obtain

$$\begin{aligned}&\text{Rank}\left(\overline{\boldsymbol{W}}^*_{b,a}\right)\\ =&\text{Rank}\left(\boldsymbol{C}^*_{b,a}\overline{\boldsymbol{W}}^*_{b,a}\right) = \text{Rank}\left(\overline{\boldsymbol{H}}_{b,a}\overline{\boldsymbol{W}}^*_{b,a}\right) \quad (24)\\ \leq& \min\left\{\text{Rank}\left(\overline{\boldsymbol{X}}^{\text{H}}_{a,b}\right), \text{Rank}\left(\boldsymbol{H}_{a,b}\right), \text{Rank}\left(\overline{\boldsymbol{X}}_{a,b}\right)\right\} = 1.\end{aligned}$$

If the matrix $\boldsymbol{C}^*_{b,a}$ is rank deficient, we can obtain a rank-one matrix $\widetilde{\boldsymbol{W}}^*_{b,a}$ via the arguments in Appendix B of [5] as

$$\widetilde{\boldsymbol{W}}^*_{b,a} = \overline{\boldsymbol{W}}^*_{b,a} - \Delta\overline{\boldsymbol{W}}^*_{b,a} \quad (25)$$

where $\Delta\overline{\boldsymbol{W}}^*_{b,a} = \sum_{k=1}^{N_b - r_b - 1}\varsigma_k\boldsymbol{\xi}_k\boldsymbol{\xi}_k^{\text{H}}$. The vector $\boldsymbol{\xi}_k$ is the $k$-th column of the null space of the matrix $\boldsymbol{C}^*_{b,a}$ and $\varsigma_k \geq 0$, $k = 1, 2, \ldots, N_b - r_b - 1$ with $r_b$ as the rank of the matrix $\boldsymbol{C}^*_{b,a}$. Similarly, when the matrix $\boldsymbol{C}^*_{a,b}$ is rank deficient, we obtain the rank-one matrix $\widetilde{\boldsymbol{W}}^*_{a,b}$ as

$$\widetilde{\boldsymbol{W}}^*_{a,b} = \overline{\boldsymbol{W}}^*_{a,b} - \Delta\overline{\boldsymbol{W}}^*_{a,b} \quad (26)$$

where $\Delta\overline{\boldsymbol{W}}^*_{a,b} = \sum_{k=1}^{N_a - r_a - 1}\zeta_k\boldsymbol{\pi}_k\boldsymbol{\pi}_k^{\text{H}}$. The vector $\boldsymbol{\pi}_k$ is the $k$th column of the null space of the matrix $\boldsymbol{C}^*_{a,b}$ and $\zeta_k \geq 0$, $k = 1, 2, \ldots, N_a - r_a - 1$ with $r_a$ as the rank of the matrix $\boldsymbol{C}^*_{a,b}$. In order to guarantee the constraints in (16b)-(16h), we modify the matrix $\overline{\boldsymbol{V}}^*$ into the matrix $\widetilde{\boldsymbol{V}}^*$ as

$$\widetilde{\boldsymbol{V}}^* = \overline{\boldsymbol{V}}^* + \overline{\boldsymbol{Y}}^{\text{H}}\begin{bmatrix}\overline{\boldsymbol{X}}_{a,b}\Delta\overline{\boldsymbol{W}}^*_{a,b}\overline{\boldsymbol{X}}^{\text{H}}_{a,b} & 0\\ 0 & \overline{\boldsymbol{X}}_{b,a}\Delta\overline{\boldsymbol{W}}^*_{b,a}\overline{\boldsymbol{X}}^{\text{H}}_{b,a}\end{bmatrix}\overline{\boldsymbol{Y}}. \quad (27)$$